\documentclass[aps,pre,superscriptaddress,twocolumn,showpacs,floatfix]{revtex4}
\usepackage{graphicx}
\usepackage{dcolumn}
\usepackage{bm}

\begin{document}

\title{Darwinian Selection and Non-existence of Nash Equilibria.}
\author{Daniel Eriksson} 
\author{ Henrik Jeldtoft Jensen}
\email[Author to whom correspondence should be addressed:\\]
{h.jensen@ic.ac.uk} 
\homepage{http://www.ma.ic.ac.uk/~hjjens/}
\affiliation{Department of Mathematics, Imperial College, 
180 Queen's Gate, London SW7 2BZ, U.K.}


\begin{abstract}
We study selection acting on phenotype in a collection of agents playing local games 
lacking Nash equilibria. After each cycle one of the agents losing most games is replaced 
by a new agent with new random strategy and game partner. The network generated can 
be considered critical in the sense that the lifetimes of the agents is power law 
distributed. The longest surviving agents are those with the lowest absolute score 
per time step. The emergent ecology is characterized by a broad range of behaviors. 
Nevertheless, the agents tend to be similar to their opponents in terms of performance. 
\end{abstract}

\pacs{87.10.+4, 87.23.-n, 02.50.Le, 05.65.+b}

\maketitle

It has been argued that biological evolution is driven by a combination of natural 
selection and self-organization \cite{Kauffman,Smith}.  Mutations act at random at 
the level of genotype while selection acts at the phenotype. Darwinian selection is 
expected to occur because some phenotypes are more viable than others \cite{Darwin}. 
The viability or fitness of a given phenotype is, of course, not an absolute quantity 
but depends on context and the environment the phenotype is exposed to \cite{Pagie}. 
The environment (physical as well as biotic) is to a large extent 
produced by co-existing phenotypes, and hence the environment and the corresponding 
fitness are self-organized emergent properties of the ecology. 
Moreover, fitness or vigor is necessarily a relative quantity. An organism can become 
so vigorous that it removes its own foundation of existence; a balance very relevant e.g. 
to host-parasite systems. For organisms to be able to coexist for extended periods of 
time they need to develop phenotypes which are in a kind of restrained poise.

Here we analyze Darwinian selection acting on phenotype in a model consisting of an 
adaptive network of agents of different strategies mutually competing in local zero-sum 
games lacking Nash equilibria. We let selection act by removing the agents with the 
smallest score (most negative) after a round of games. Thus the viability of an agent 
is a function of his phenotypical behavior in a specific environment. When the local 
games are of {\it opposing interest} with only a single winner we find that the agents 
organize into a system where agents are homogeneous in terms of success but heterogeneous 
in terms of activity. The agents achieve longevity by organizing themselves in a way 
that minimize the absolute value of their score. By this strategy they avoid to be 
amongst the worst performers themselves; moreover low absolute score allows the 
agents to retain well tuned partners by not forcing many lost games on to their 
opponents, which would lead to the elimination of the partner. 

Our model is closely related to previous network models. Kauffman introduced Boolean 
networks to study the relationship between evolution and self-organization \cite{Kauffman}.  
Collective adaptive agents striving to be among the minority were studied in mean-field 
models by Arthur \cite{Arthur} and Challet and Zhang \cite{Challet}.
Aspects of Darwinian evolution were added to the minority game by 
Zhang \cite{Zhang}.  
This line of approach were further developed by Paczuski, Bassler and Corral (PBC) 
who considered agents playing the minority 
game \cite{Paczuski}. The goal of agents is to be among the {\it global} 
minority though they are linked only to a subset of all the agents. The 
strategy of the worst performing agent is randomly changed after each cycle 
while the links of the network remain unchanged \cite{Paczuski}. 

Here we modify the model introduced by PBC to a Local Darwinian Network model (LDN) in 
which we allow agents to play competitive {\it local} games and to adapt their connections 
as well as their strategies. Viewed from the perspective of evolutionary ecology it 
appears to be more natural to consider agents performing local games 
(organisms or species entangled in a web of mutual competitions). The renewal of the 
properties of the worst performing agent should be thought of as representing either 
the ``mutation'' of an individual or, if agents are thought of as representing species, 
as a species being superseded by an invading species with different affinities. 
In either case it is natural to update the set of interaction links. Further,
from a statistical mechanics or complex systems perspective we expect systems coupled 
globally to more readily enter a critical state, hence it is of interest to explore the 
criticality of purely local systems. 

The LDN net is critical for $K=2$ as is the original Kauffman net \cite{Kauffman}. 
The PBC net is critical for $K=3$ \cite{Paczuski}. Here $K$ is the number of independent 
arguments of the agents' Boolean strategy functions. 

The homogeneity parameter $P$ of an agent measures the fraction of 0's and 
1's -- whichever is in majority -- in the Boolean output assigned to the $K^2$ 
distinct input states of the agent. For $K=2$, the homogeneity parameter is constrained 
to one of the values $1/2$, $3/4$ or $1$ where in the last case the agent will not 
switch output signal at all. In contrast to PBC, we fix the homogeneity parameter of 
each agent to $3/4$ ensuring that the network is logically entirely homogeneous with 
respect to the output of the agents. It follows that any discrepancy in the switching 
behavior of two competing agents is of dynamical origin.

{\it The model} -- Consider N agents, each assigned a Boolean signal $S_i = 0$ or $1$,
 for $i=1,..., N$. The dynamics consists of two types of updates: a) 
the agents playing rounds of games; b) and the act of Darwinian selection/mutation. 

a) {\bf The games} are performed in the following way. Each agent is defined as aggressor 
({\bf \cal A}) of one other randomly assigned agent (thus agents participate on average 
in two games) who acts as opponent ({\bf \cal O}) in a zero sum game. In each time step 
{\bf \cal A} and {\bf \cal O} compare their Boolean signals $S_{\bf \cal A}$ 
and $S_{\bf \cal O}$. If $S_{\bf \cal A}=S_{\bf \cal O}$  the aggressor {\bf \cal A} 
scores +1 and the opponent scores --1. If $S_{\bf \cal A}\neq S_{\bf \cal O}$  
the aggressor {\bf \cal A} scores --1 and the opponent scores +1.  The game is 
identical to the simple "coin guessing" game where one of two persons is to pick the
 hand -- left or right - that is holding the coin. The game has the property that, 
when considering only pure (non-randomized) strategies, 
there exists no Nash equilibrium. No Nash equilibrium exists in the
sense that it will always be possible for either player to benefit by changing his 
strategy if the opponent sticks to his strategy. In our simulations, two agents are 
never allowed to act as mutual aggressors in which case both agents could receive a 
zero score. Neither do we allow agents to act as aggressor for themselves. The fact 
that agents cannot collaborate in order to achieve neutral (zero) scores will be 
referred to as an opposing interests property of the network. 

The signals used for the zero-sum games are generated in the following way. The output 
signal at a given time step $S_i(t)$ of agent $i$ is determined 
deterministically from the output at the previous time step from $K$ other {\em source}
agents $S_{i_1}(t-1),S_{i_2}(t-1),...,S_{i_K}(t-1)$ 
\begin{equation}
S_i(t) = f_i( S_{i_1}(t-1), S_{i_2}(t-1), ... , S_{i_K}(t-1) ) ,   
\label{e_q_m}
\end{equation}
where $f_i$ is a randomly chosen Boolean function associated with agent $i$. 
For $K=2$ the function is such that for exactly three of the functions four input 
configurations ($\{0,0\}$, $\{0,1\}$, $\{1,0\}$, $\{1,1\}$) the same Boolean output is 
assigned, whereas the signal is different for the fourth configuration. This 
ensures $P=3/4$.

One can think of a specific choice for the functional form of $f_i$ as the genotype 
of agent number $i$. The $K$ source-agents associated with $i$ can include the opponent 
of $i$ but will not include agent $i$. The source-agents and the opponents
are to be thought of as the 
environment of an agent.

For fixed assignments of source-agents as well as aggressor and opponents 
Eq. {\ref{e_q_m} is now parallel updated $\forall i$. Since the state space of the 
net is finite and the dynamics is deterministic, periodic orbits will always be reached 
(though the maximum length of a period is huge $2^N$). We measure the average score per 
time step of the agents over a round of games consisting of either simulating the 
transient plus one periodic cycle or by performing $10^4$ parallel updates which ever 
is the shortest. We denote this sequence of deterministic updates a {\em test cycle}. 
The score gained by an agent characterizes the success of his phenotypical behavior. 

b) {\bf The Darwinian update} is done in the following way. After testing the performance 
of the agents, the worst performing agent is replaced by a new agent with a new Boolean 
function (i.e., genotype) chosen completely at random. (In case the worst performance 
is shared by more then one agent, then one of them is chosen at random for replacement.)
The new agent is randomly assigned an opponent and a set of $K$ source-agents. 
Intuitively, this means that agents belonging to ecological niches that no longer 
exist are removed from the system, whereas new candidate niches are sampled 
fortuitously by agents who themselves have random properties. However, the wiring of 
those agents that attack an agent being replaced is not changed. These agents now attack 
the new agent as we proceed with a new round of games, or test cycle, as described 
under a). 

In this letter, we use a Non-Darwinian version of the model as benchmark allowing us 
to isolate the effect of the Darwinian updating. In the Non-Darwinian model, at the 
end of each test cycle, we randomly pick an agent for replacement. Agents are 
in this situation not punished for performing poorly and are all subject to the 
same probability of being replaced.

{\it Results} - The initial configuration consists of randomly assigned input links, 
Boolean functions and game partners. As the sequence of test cycles and Darwinian 
replacements are repeated the model organizes gradually into a stationary state.
The critical properties of the network are indicated in Fig. \ref{periods} where 
we exhibit the probability of the length of transients $p(t)$ and of the length 
of periodic attractors $p(a)$ respectively. The distributions do exhibit scale free 
power law like behavior, though an accurate determination of the precise functional 
form (and precise values of the exponents) is unfortunately not possible. The behavior 
of $p(t)$ for $K=2$ for large $N$ appears to be consistent with $p(t)\sim t^{-\alpha}$ 
for $\alpha\simeq 1.4$. The behavior of $p(a)$ for $K=2$ for large $N$ appears to 
consistent with $p(a)\sim a^{-\beta}$ for $\beta\simeq 1.2$. The behavior of the 
Non-Darwinian version of the network is similar as shown in the figures. 
Hence the Darwinian move is not crucial for the existence of the power
law like behavior of $p(t)$ and $p(a)$. Nevertheless we shall see below
that the Darwinian selective move has a significant effect on the characteristics
of the agents. A transient and 
corresponding first occurrence of periodic attractor make up a test cycle. We find 
that the distribution of test cycles also appears to be power law like. Notice that in 
order to find the largest observations in Fig. \ref{periods} 
we iterated the network more than the maximum number of updates of a test cycle and 
then reentered the state at $10^4$ updates to perform the Darwinian update.

We define the lifetime $l$ of an agent as the number of updates survived at the moment 
the agent is removed from the system. As exhibited in Fig. \ref{lifetime} we find that 
under Darwinian updating the network develops what appears to be a power 
law like distribution of lifetimes $p(l)$ with increasing system size, consistent with 
$p(l)\sim l^{-\gamma}$ for $\gamma\simeq 0.9$. Notably the magnitude of the power 
law exponent is smaller than one. Since test cycles are numerically truncated at a 
finite length of $10^4$ updates, the distribution of lifetimes is somewhat distorted 
for lengths longer than the numerical cutoff. Because of logarithmic binning the effect 
is not visible in Fig. \ref{lifetime}.  We find that for all considered values of 
$N=64, 128$ and 1024 the  lifetime distribution for the Non-Darwinian model (the circles
in Fig. \ref{lifetime}) are approximately linear as function of the {\em logarithm} 
of the argument, i.e. not at all described by a power law.

\begin{figure}
\scalebox{0.5}[0.35]{\includegraphics{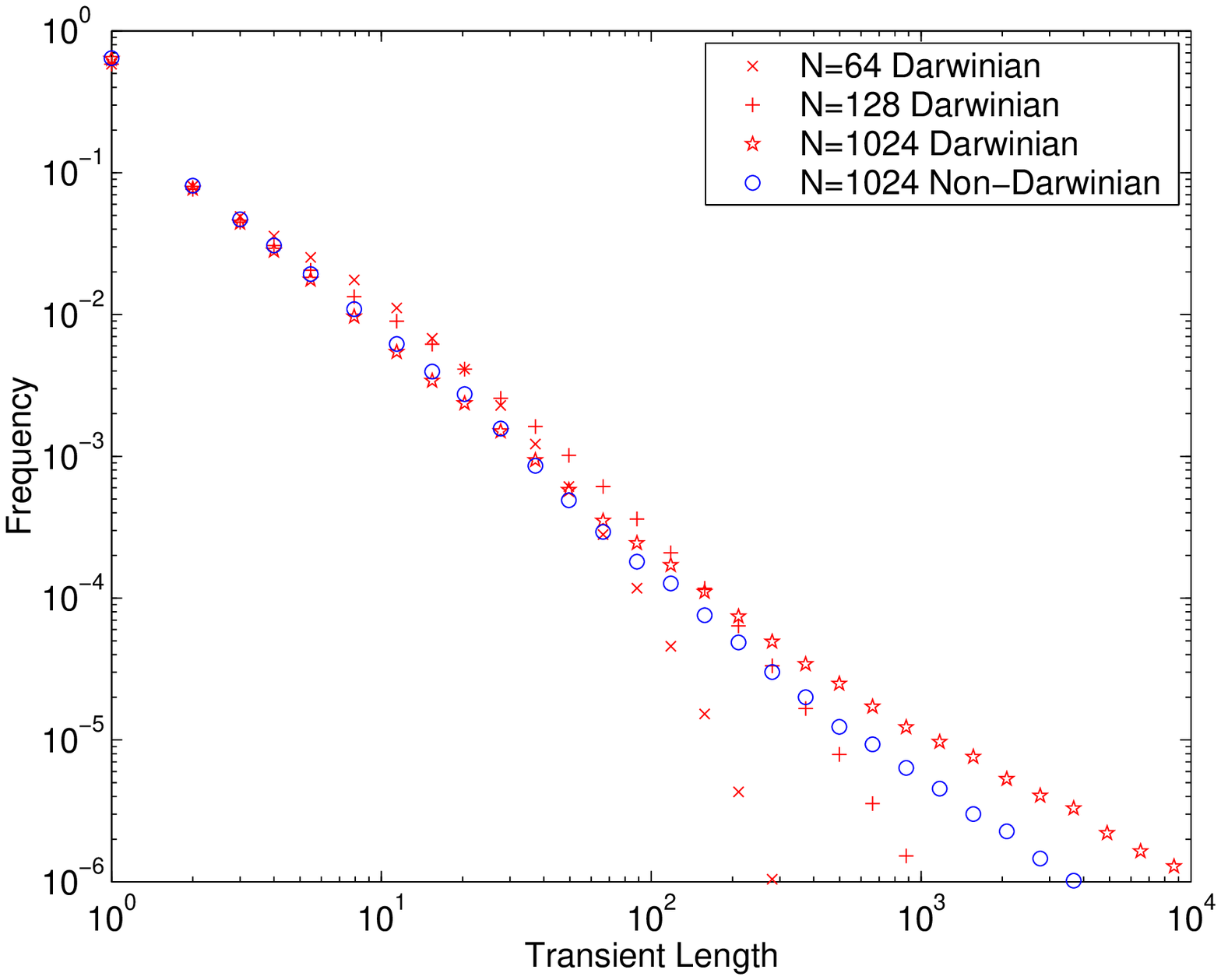}}
\scalebox{0.5}[0.35]{\includegraphics{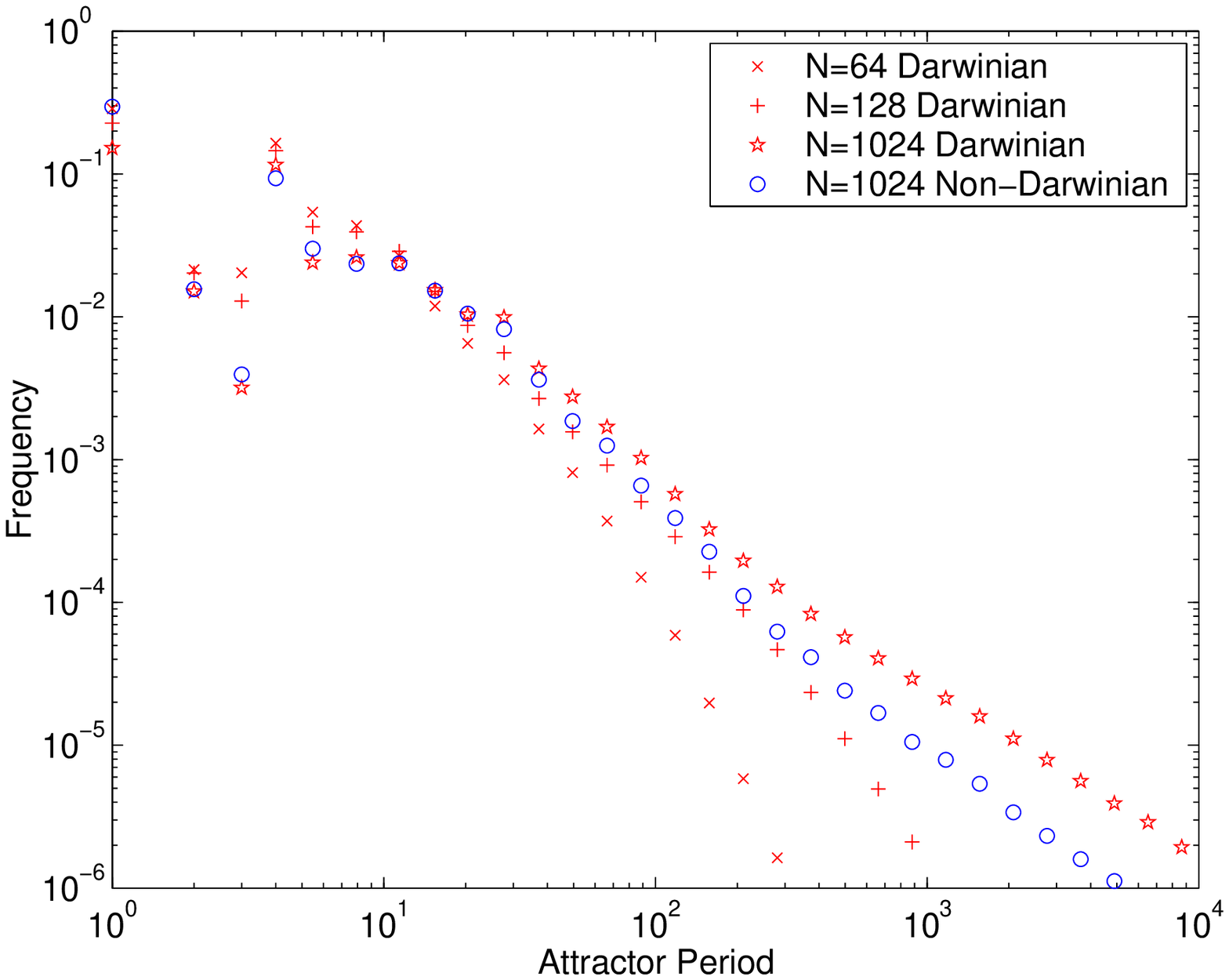}}
\caption{The probability of the length of transients
(top curve) and attractors (bottom curve) for different systems sizes $N$. 
\label{periods}}
\end{figure}  

\begin{figure}
\scalebox{0.5}[0.35]{\includegraphics{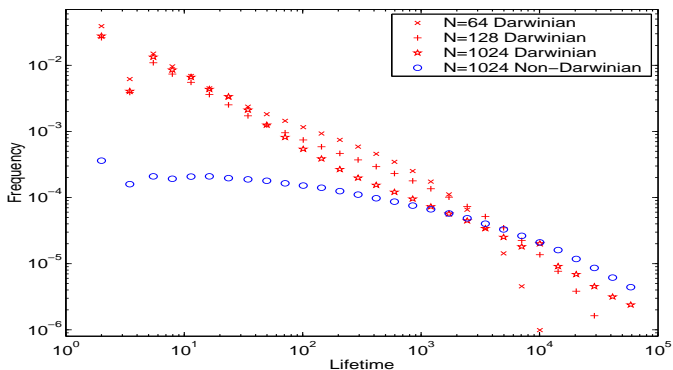}}
\caption{The probability of the number of updates agents survive
for different system sizes $N$. 
\label{lifetime}}
\end{figure} 

Next we address the behavior of the agents in the stationary critical state. We denote 
by the lifetime {\em yield} of an agent the score averaged over the lifetime of the 
agent. The yield of an agent characterizes the efficiency of the agent's phenotypical 
behavior. Fig. \ref{lifetime_yield}  shows that longevity is directly related to 
near zero values of the agent's lifetime yield.  Thus the selective pressure on the 
agents in the model leads to a finely tuned ecology, where agents tend to avoid 
winning too much, which might induce annihilation of their partners. Instead agents 
manage to keep their partners and live long by entering into near neutral patterns. 

\begin{figure}
\scalebox{0.5}[0.35]{\includegraphics{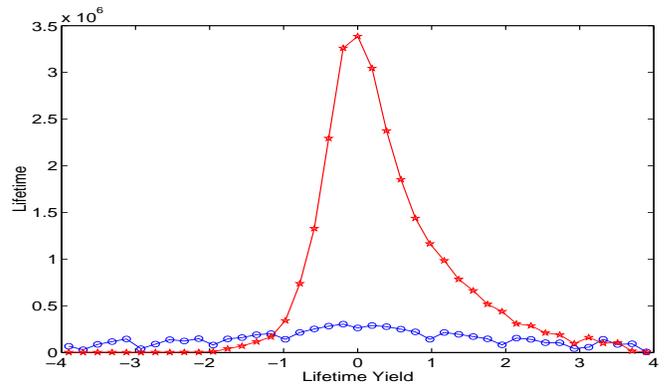}}
\caption{The correlation between lifetime and lifetime yield for the Darwinian model (stars) and the Non-Darwinian model (circles). $N=1024$.} 
\label{lifetime_yield}
\end{figure} 

Fig. \ref{abs_ave_score} is a histogram of yield accumulated over entire
test cicles of the Darwinian and benchmark ecologies respectively. The accumulated yield 
of an agent is the average score of the agent over the length of his life so far. 
(Notice that agents were not born at the same time.) The standard deviation 
$\delta Y$ of the distribution of accumulated yield is a measure of the equality 
in success in the network. We find a smaller $\delta Y$ in the Darwinian ecology 
for all considered system sizes $N=128$, $256$ and $1024$. 

Indeed, one would expect that removal of the worst performing agents should reduce the 
abundance of poorly performing agents with large negative average scores. But we 
observe that the abundance of well performing agents with large positive scores is 
also somewhat reduced. In the Darwinian model agents are thus clustered around the 
zero score. Notice that given opposing interests of agents, the collected environment 
of an agent receives the same score as the agent but with opposite sign. In the 
Darwinian model, agents are therefore similar to their respective environments in 
terms of success. An interesting property of e.g. the Minority Game (MG)
model is that any 
favorable configuration for a particular agent is inherently unstable as other agents 
act to reverse the situation \cite{Challet}. A similar situation holds for LDN as well 
but unlike MG the environments of the agents are heterogeneous. 

Let us now describe in more detail how the agents manage to achieve low absolute scores 
and thereby survive many test cycles. For the benchmark and Darwinian model respectively, 
we have investigated the switching activity of the agents. 
By a switch we mean the change from sending signal $0$ to $1$ or vice versa. By a distinct 
switching activity we mean a certain observed number of switches of one or more agents 
over a test cycle. The number $W(N, t+a)$ of observed distinct activities
cannot be larger than the number of agents in the ecology $N$ nor the length of the 
test cycle $(t+a)$. We have investigated the dependence of $W$ on $(t+a)$ for system 
sizes $N=128$, $256$ and $1024$. We find that for all cycle lengths $2 \leq (t+a) < 10^4$, 
the Darwinian ecology exhibits a larger ratio $W/(t+a)$. This may suggest that the 
Darwinian updating of the network produces a greater variety of phenotypes.

\begin{figure}
\scalebox{0.5}[0.35]{\includegraphics{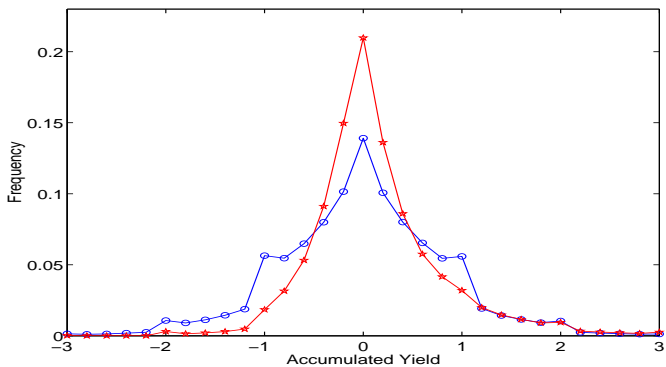}}
\caption{ Histogram of accumulated yields for the Darwinian model (stars)
and the Non-Darwinian model (circles). $N=1024$.
\label{abs_ave_score}}
\end{figure} 

We now discuss the robustness of the model. Increasing the number of opponents per 
agent does not appear to change the test cycle distribution. We do not observe 
criticality in case the input wiring of the agents is unchanged as 
the agents are replaced. We point out that, in general, testing the stationarity 
of the model is numerically challenging. One can consider an alternative version of 
the network where the homogeneity parameter is not fixed. In this situation, PBC find 
that when agents compete globally, the homogeneity parameter self-organizes to a 
distinct value as the network enters a critical state \cite{Paczuski}. In contrast, 
for the LDN model we find that the homogeneity parameter decreases from the initial 
unbiased value making the transients and periodic orbits very long,
tough still, perhaps, power law distributed. This version of the model is
difficult to handle numerically.

As pointed out by PBC, an interesting question is -- what games lead to complex, 
scale free states under Darwinian selection? We speculate that the game should 
inherently prevent agents from achieving mutual gain while allowing the losing 
agents to improve their unfavorable situation by a change of behavior. 

We suggest that the "coin-guessing" game of the LDN model is a suitable representation 
of such situations where agents tend to become equal in strength to their opponents 
implying their exact "task" becomes irrelevant. For example, the game is similar to 
many situations in the business world, e.g. the trade of a stock involving one seller 
and one buyer. From a dynamical point of view, it is significant that either the buyer 
or seller in a trade will necessarily become a winner and the other agent a loser as 
the price of the stock diverges from its value when the deal was made. (If the price 
of the stock goes up the seller would have been better off on the other side of the trade). As in the LDN model, the determinant of who is the winner is of complex dynamical origin. 

The score distribution of the Darwinian model might be compared to e.g. the 
distribution of performances of fund managers. Interestingly, extensive 
investigations have given little evidence that fund managers can consistently 
beat relevant market indexes (see p. 368 in \cite{Bodie}). Similarly, the LDN 
model has the property that successful agents tend to be short-lived. 

We have presented a Local Darwinian Network model which exhibits critical behavior 
without global interactions. The model demonstrates that -- given that agents act 
under opposing interest with their respective environments -- Darwinian phenotypical 
selection of the worst performer together with genotype mutations can lead to an ecology 
where the distribution of lifetimes lacks a characteristic scale. The agents of the 
emergent ecology are heterogeneous with respect to their activity but in terms of 
success they are nearly equal to their respective environments. Longevity is achieved 
by adapting to a near neutral yield which ensures the stability of the fellow partners.

Helpful initial inspiration from M. Paczuski, P. Bak and K. Nagel are gratefully 
acknowledged by DE.

\bibliography{dan}
\end{document}